\documentclass[aps,prl,epsf,twocolumn]{revtex4}
\usepackage{graphicx}

\usepackage{bm}

\usepackage[intlimits]{amsmath} 

\newcommand{\bra}[1]{\langle #1|}
\newcommand{\ket}[1]{|#1\rangle}

\newcommand{\eqnref}[1]{Eq.~(\ref{#1})} 
\newcommand{\fref}[1]{Fig.~\ref{#1}}    
\newcommand{\sigmdag}{{\hat {\sigma}^\dag}}
\newcommand{\sigm}{{\hat {\sigma}}}
\newcommand{\adag}{{\hat{a}^\dag}}
\newcommand{\bdag}{{\hat{b}^\dag}}
\newcommand{\Hint}{\hat H_{\textrm{int}}}

\newcommand{\piplus}{{\hat\pi}^{+}}
\newcommand{\piminus}{\hat\pi^-}
\newcommand{\piz}{\hat\pi_z}
\newcommand{\hc}{\mathrm{H.c.}}
\newcommand{\St}{\hat {S}}
\newcommand{\Stdag}{\hat {S}^{\dag}}
\newcommand{\arctanh}{\operatorname{arctanh}}
\newcommand{\abs}[1]{\left|#1\right|}
\newcommand{\OmBog}{{\Omega_\textrm{b}}}
\newcommand{\hrho}{{\hat \rho}}
\newcommand{\rate}{r_{\textrm{at}}}
\newcommand*{\pdiff}[3][]{\frac{\partial^{#1}{#2}}{\partial^{}{#3}{}^{#1}}}

\newcommand{\ave}[1]{\langle#1\rangle}

\begin{document}

\title{Generation of EPR-entangled radiation through an atomic reservoir}
\author{Susanne Pielawa,$^{1,2}$ Giovanna Morigi,$^1$ David Vitali,$^3$ and Luiz Davidovich $^4$}
\affiliation{
$^1$ Departament de Fisica, Universitat Aut\`onoma de Barcelona, 08193 Bellaterra, Spain\\
$^2$ ICFO -- Institut de Ci\`encies Fot\`oniques, 08860 Castelldefels (Barcelona), Spain\\
$^3$ Dipartimento di Fisica, Universit\`a di Camerino, 62032 Camerino, Italy\\
$^4$ Instituto de F\'isica, Universidade Federal do Rio de
Janeiro, 21941-972 Rio de Janeiro, Brazil } \date{\today}
\begin{abstract} We propose a scheme for generating two-mode
squeezing in high-$Q$ resonators using a beam of atoms with random
arrival times, which acts as a reservoir for the field. The scheme
is based on four-wave mixing processes leading to emission into
two cavity modes, which are resonant with the Rabi sidebands of
the atomic dipole transition, driven by a saturating classical
field. At steady state the cavity modes are in an
Einstein-Podolski-Rosen (EPR) state, whose degree of entanglement
is controlled by the intensity and the frequency of the transverse
field. This scheme is robust against stochastic fluctuations in
the atomic beam, does not require atomic detection nor velocity
selection, and can be realized by presently available experimental
setups with microwave resonators. \end{abstract} \maketitle

A single atom is a text-book example of a nonclassical light
source. Seminal experiments have shown the sudden nature of
spontaneous emission events~\cite{Q:Jumps}, the quantum properties
of the emitted photons~\cite{Antibunching, Hong-Ou-Mandel}, and
entanglement between the atom and its emitted
photon~\cite{Monroe04&Weinfurter2005}. The high degree of control
on the dynamics of the interaction between single atoms and the
electromagnetic field has allowed the access to novel regimes,
making single atoms in resonators promising candidates, together
with atomic ensembles~\cite{Ensembles}, for implementing
interfaces for quantum networks~\cite{Lukin03}. In optical
cavities, for instance, lasing at the single-atom
level~\cite{An94+Kimble-atomlaser} and controlled single-photon
generation~\cite{Kuhn02+Kimble+Keller04} have been demonstrated.
In the microwave domain remarkable milestones have been achieved
thanks to the extremely stable resonators that are available in
this frequency
domain~\cite{Haroche-Colloquium,CavityQED-Walther,ultra}. Some
paradigmatic experiments are the preparation and measurement of
non-classical states of the microwave
field~\cite{Haroche-Colloquium,CavityQED-Walther}, the
experimental characterization of loss of coherence of the quantum
field~\cite{Decoherence} and of the transition from quantum to
classical dynamics~\cite{Quantum-Classical}, and the quantum
non-demolition measurement of the number of photons of the cavity
field~\cite{QND-Nature}. In the weak-coupling limit, the atomic
beam may act as a reservoir for the cavity mode, leading to
thermalization between the beam and the
resonator~\cite{master-eq}. The opposite regime, with the
resonator field saturating the atomic transition, leads, for
velocity-selected atoms, to sub-Poissonian fields, and, in the
weak-dissipation limit, to trapping
states~\cite{Meystre,WaltherTrapping}. Methods for preparation of
an arbitrary single-mode quantum state of the electromagnetic
field in a resonator, involving resonant interaction with a
well-controlled sequence of atoms, without the need of atomic
detection, were proposed in~\cite{Zoller,Law,Wellens00}. Methods
requiring atomic-state measurement, on the other hand, rely on
high-efficiency detectors, which are lacking in present
experiments.

\begin{figure}[b]
\includegraphics[width=0.4\textwidth]{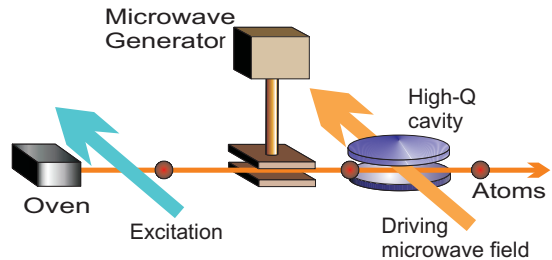}
\caption{Setup of the system. Atoms from a beam are prepared in a
coherent superposition of two Rydberg states $|g\rangle$ and
$|e\rangle$ by a combination of laser and microwave fields.
The atoms have random arrival times, and a low pumping rate warrants that at most one atom
is inside the resonator at a time~\cite{Haroche-Colloquium, CavityQED-Walther}.
While in the cavity, the dipole transition $\ket g \rightarrow \ket e$ is saturated by a transverse microwave
field, thereby pumping on resonance two non-degenerate modes of
the resonator,  which are led asymptotically to a two-mode squeezed state.}
\label{Fig:exp-setup} \end{figure}

In this Letter, we propose a method for preparing quantum states
of the electromagnetic field based on quantum reservoir engineering. This
method, so far applied to trapped ions~\cite{reservoir}, is here
implemented in a typical setup of microwave cavity QED as in
Fig.~\ref{Fig:exp-setup}, where the resonator is pumped
by a beam of atoms with random arrival times, and needs neither
atomic detection, nor detailed control of the sequence of
atoms: We show that, by suitably preparing the initial state of
the incoming atoms, two-mode squeezing, i.e., Einstein-Podolski-Rosen
correlations~\cite{EPR}, are established between the cavity modes
at steady state.

The underlying mechanism is four-wave mixing, where emission into
the cavity modes is enhanced by resonant coupling with the Rabi
sidebands induced by a classical field that saturates the atomic
transition~\cite{Agarwal}, with the creation of EPR-correlations
being enforced by the initial quantum state of the injected atoms. Contrary to typical
setups based on optical parametric amplifiers, the atoms pump the
resonator through resonant single-photon processes. Differing
from~\cite{Wellens00,Law}, the atoms do not need to be initially correlated
nor their number has to be controlled: The atomic beam acts as a
reservoir, where the atoms interact with the field one at a
time~\cite{Scarani02}, and that pulls the field into the desired
state. The degree of entanglement is controlled by the intensity
of the laser and its detuning from the atomic transition. Many
photons per mode can be achieved in accessible experimental
regime. Control on atomic velocity (interaction time) and atomic
detection are not required. The scheme is robust against
stochastic fluctuations in the atomic beam. This method may
constitute an important step towards the implementation of quantum
networking with continuous variables~\cite{Braustein04} in the
microwave regime.

\begin{figure}[t] \includegraphics[width=0.4\textwidth]{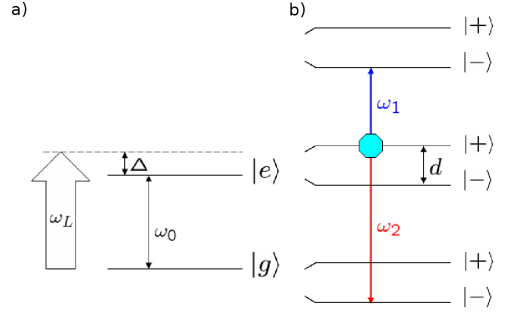}
\caption{(a) Dipole transition $|g\rangle\to |e\rangle$ at
frequency $\omega_0$ coupling to the classical field at frequency
$\omega_L$. (b) Resulting ladder of the semiclassical dressed
states $\ket \pm$. A transition $\ket + \to \ket -$ is accompanied
by absorption (emission) of a photon of frequency $\omega_1$
($\omega_2$), and viceversa -- see text.} \label{Fig:ladder}
\end{figure}

The basic system is sketched in~\fref{Fig:exp-setup}. Prior to the
interaction region, the atoms are prepared in a coherent
superposition of two Rydberg states $\ket g$ and $\ket e$
connected by a dipole transition. For an open-cavity
geometry~\cite{Haroche-Colloquium}, an electric potential between
the two mirrors removes through Stark shifting the degeneracy of
circular Rydberg states. Inside the resonator a classical field
saturates the dipole transition, thereby pumping on resonance two
non-degenerate modes of the resonator at frequencies $\omega_1$
and $\omega_2$, as sketched in Fig.~\ref{Fig:ladder}.  The
corresponding Hamiltonian is similar to the one describing
four-wave mixing as in \cite{Agarwal}, and has the form \begin{eqnarray}
\hat H&=&\hbar\omega_0\hat\sigma^\dagger\hat\sigma+
\hbar\Omega\left(e^{-i\omega_Lt}\hat\sigma^\dagger+e^{i\omega_Lt}\hat\sigma\right)\nonumber\\
&+&\sum_\lambda\left[\hbar\omega_\lambda\hat a^\dagger_\lambda
\hat a_\lambda+\hbar g_\lambda(\hat
a_\lambda\hat\sigma^\dagger+\hat
a^\dagger_\lambda\hat\sigma)\right]\,, \end{eqnarray} where
$\omega_0$ is the transition frequency, $\hat a_\lambda$ and
$\adag_\lambda$  are photon annihilation and creation operators
for the mode with frequency $\omega_{\lambda}$ ($\lambda = 1, 2$),
$g_\lambda$ are the coupling constants between the two-level atom
and each cavity mode, and $\sigmdag = \ket e \bra g$, $\sigm =
\ket g \bra e$ are the atomic raising and lowering
operators~\cite{polarizations}. The time-dependent term describes
the coupling, with strength $\Omega$, between the dipole and the
external classical field at frequency $\omega_L$.

The transformation $\hat
U=\exp\left[i\omega_Lt\left(\hat\sigma^\dagger\hat\sigma+\sum_\lambda\hat
a^\dagger\hat a\right)\right]$ takes the above Hamiltonian into
the rotating-frame form \begin{eqnarray} \hat H^{RF}=\hat {\cal
H}_0- \sum_{\lambda}\hbar\delta_\lambda\adag_\lambda\hat a_\lambda
+\sum_{\lambda}  \hbar g_\lambda \left( \sigmdag \hat a_\lambda +
\sigm \adag_\lambda \right), \label{eqn:Hamiltonian-laser-frame}
\end{eqnarray} where $\delta_{\lambda}=\omega_L-\omega_{\lambda}$,
and $\hat {\cal H}_0=- \hbar \Delta \sigmdag\sigm + \hbar\Omega
\left(\sigmdag + \sigm\right)$ describes the coupling between
dipole and classical field, with detuning $\Delta = \omega_L -
\omega_0$, see~\fref{Fig:ladder}(a). We assume the coupling to the
driving field to be much stronger than the coupling to the cavity
modes, $|\Omega| \gg |g_\lambda|$, and express
Hamiltonian~(\ref{eqn:Hamiltonian-laser-frame}) in the basis of
eigenstates $\ket{\pm}$ of $\hat {\cal H}_0$, with $\hat {\cal
H}_0\ket{\pm}=-\hbar(\Delta\mp d)/2\ket{\pm}$ and $d =
\sqrt{\Delta^2 + 4\Omega^2}$. These are the semiclassical dressed
states, with $\ket{+}=\sin\theta\ket g + \cos\theta\ket e$, $\ket
{-} = \cos\theta\ket g - \sin\theta\ket e$, and $\tan\theta =
2|\Omega|/(d-\Delta)$. The corresponding energy levels are shown
in ~\fref{Fig:ladder}(b). We denote the raising and lowering
operators in the new basis by $\piplus = \ket +\bra -$ and
$\piminus = \ket -\bra +$, with $\piz = \ket +\bra + - \ket -\bra
-$, and write $\hat H^{RF} = \hat H_0 + \Hint$, with ${\hat H_0 =
\hbar d\piz/2 -\hbar\sum_{\lambda}\delta_{\lambda}
\adag_{\lambda}\hat a_{\lambda}}$ and \begin{eqnarray} \label{eqn:
Hint-total} &&\Hint = \sum_{\lambda}\hbar g_\lambda \left[
\piz\left(\hat a_\lambda +\adag_\lambda\right)\cos\theta\sin\theta
\right.\\&& \left.+\left(\piplus\hat a_\lambda
+\adag_\lambda\piminus\right)\cos^2\theta - \left(\piminus\hat
a_\lambda+\piplus\adag_\lambda\right)\sin^2\theta\right].
\nonumber \end{eqnarray} If $|g_\lambda|\ll d$ we can choose which
processes are resonant, and thus relevant for the dynamics, by
changing the values of $\delta_1, \delta_2$, and $d$. One is thus
able to generate a diversity of dynamical processes. For instance,
generation of Schr\"odinger cat states of a single-mode resonator,
as discussed in Ref.~\cite{Solano}, is recovered from $\hat
H^{RF}$ by setting $d=-\delta_1$ and $|\delta_2|\gg
d,|\delta_1|,g_{\lambda}$, hence obtaining $\hat
H^{RF}\approx\hbar g_1  \piz\left(\hat a_1 +\adag_1\right)$. The
Hamiltonian of the optical parametric amplifier is obtained
from $\hat H$ by selecting resonant two-photon processes and
off-resonant single-photon transitions. The latter, however,
produce a.c.-Stark shifts which limit the efficiency of such
scheme. Instead, we propose here the generation of robust two-mode
squeezed states using {\it resonant single-photon} processes.

\begin{figure*}[ht] \includegraphics[width=0.8\textwidth]{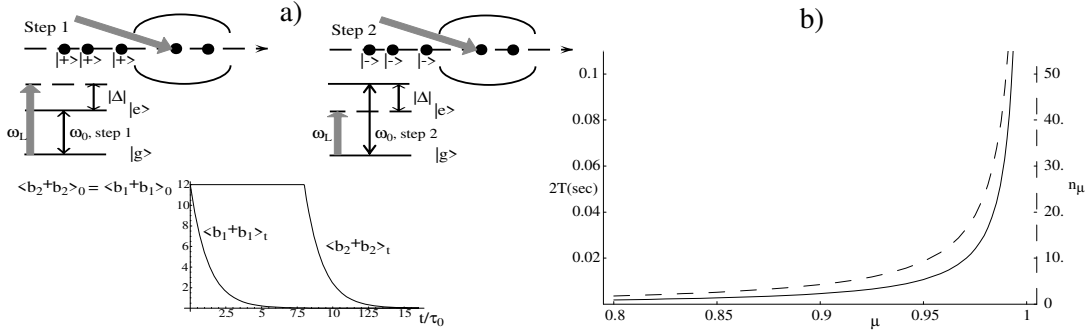}
\caption{ (a) Schematic representation of the two-step procedure.
In the plot, the expectation values of the operators $\bdag_1\hat
b_1$ and $\bdag_2\hat b_2$ are shown as a function of time, in
units of $\tau_0=\gamma^{-1}$, for $\mu=0{.}97$. After a
sufficiently long time $T$ the steady state $\ket{0,0}_b$,
corresponding to the two-mode squeezed state of the cavity modes,
is reached with fidelity $F=0.98$. (b) Average photon number
(dashed line) and total experimental time $T$, in seconds, as a
function of $\mu$.} \label{Fig:two-step} \end{figure*}

These single-photon processes are selected by setting $
\delta_1 = d$ and $\delta_2 = -d$, see~\fref{Fig:ladder}(b). If
 $|g_{\lambda}|\ll d$ we obtain from~\eqnref{eqn:
Hint-total} the effective Hamiltonian $\hat H^{RF}\approx \hat
H_{\rm eff}$, with \begin{equation} \hat H_{\rm eff}=\hat
H_0+\left[\hbar g\left(\adag_2 \cos^2\theta-\hat a_1
\sin^2\theta\right)\piminus+\hc\right]\,, \label{eqn: Hinteff}
\end{equation} where $\hat H_0 = \hbar d(\piz/2 - \adag_1\hat
a_1+\adag_2\hat a_2)$ and we have assumed $g:=g_1=g_2$. The
processes described by Eq.~(\ref{eqn: Hinteff}) are indicated by
the arrows in~\fref{Fig:ladder}(b). Using the two-mode squeezing
operator $\St(r_\mu) = \exp\left(r_\mu^*\hat a_1\hat a_2 - r_\mu
\adag_1\adag_2\right)$ we bring Hamiltonian~(\ref{eqn: Hinteff})
to the well-known Jaynes-Cummings form, $\Stdag (r_\mu)\hat H_{\rm
eff} \St(r_\mu)=\hat H_0+\hat H_{\rm int}$, with $\hat H_0=\hbar d
\left(\piz/2-\hat b^\dag_1\hat b_1+\hat b^\dag_2\hat b_2\right)$
and
\begin{eqnarray}
&&\Hint=-\hbar\OmBog \left( \hat b_1 \piminus + \hat b^\dag_1\piplus\right) ,
{\rm ~~if~~~} \Delta>0\,
\label{eqn: Hint in bog basis+}\\
&&\Hint=\hbar\OmBog\left( \hat b^\dag_2 \piminus + \hat b_2\piplus \right) ,
{\rm ~~if~~~}\Delta<0\,. \label{eqn: Hint in bog basis-}
\end{eqnarray}%
Here, $\OmBog = g \sqrt{(1-\mu)/(1+\mu)}$ with
$r_\mu=\arctanh\mu$, while the value of $\mu$ is determined by the
classical field parameters, $\mu=\tan^2\theta$ if
$\abs{\tan\theta} < 1$, else $\mu=(\tan\theta)^{-2}$ if
$\abs{\tan\theta} > 1$. The new bosonic operators $\hat b_1$,
$\hat b_2$,
are connected to $\hat{a}_1$ and $\hat{a}_2$ by the two-mode
squeezing transformation, $\hat{b}_j=\Stdag (r_\mu)\hat a_j\,
\St(r_\mu)$. The sign of the detuning $\Delta$ determines to which
of the transformed modes the two-level transition
couples~\cite{zero}. Let $\ket{n_1, n_2}_a$ and $\ket{n_1, n_2}_b$
be the eigenvectors of the number operators $\adag_j\hat a_j$ and
$\bdag_j\hat b_j$, respectively, corresponding to the eigenvalues
$n_j=0,1,2,\ldots$ ($j=1,2$). The two bases are related by the
transformation $\ket{n_1,n_2}_a=\St(r_\mu)\ket{n_1,n_2}_b$. In
particular, the vacuum state in the $b$-basis is a two-mode
squeezed state of the two cavity modes,
$\ket{0,0}_b=\Stdag(r_\mu)\ket{0,0,}_a$, with degree of squeezing
determined by $\mu$ and thus by the ratio $\abs{\Delta/\Omega}$.
For $\Delta > 0$ ($\Delta < 0$) the state $\ket{+,0,0}_b$
($\ket{-,0,0}_b$) is the ground state of the new Hamiltonian.

We now show how the cavity modes can be prepared in the two-mode
squeezed state asymptotically. This is achieved by an effective
"{\it dissipation}" process in the b-basis, implemented in a two
step procedure sketched in Fig.~\ref{Fig:two-step}(a). Step 1: One
sets $\Delta=\Delta_0>0$. The atoms enter the cavity in state
$|+\rangle$ and undergo the dynamics of Eq.~(\ref{eqn: Hint in bog
basis+}), removing in average excitations from mode $b_1$. Step 2:
Dynamics of Eq.~(\ref{eqn: Hint in bog basis-}) is selected by
setting $\Delta= -\Delta_0$. The atoms enter in the state $\ket -$
and absorb in average excitations from mode $\hat{b}_2$.

We assume the weak coupling regime, where the interaction of a
single atom with the cavity is a small
perturbation~\cite{Footnote:Master}. Let $\tau$ be the interaction
time, with $\OmBog\tau\ll 1$, and let all atoms be initially
prepared in state $\ket +$ in step~1, and in state $\ket -$ in
step~2. The differential change on the density matrix $\hrho_t$ of
the cavity during each step $j$ ($j=1,2$) is \cite{master-eq}
\begin{equation}
\left.\pdiff{\hrho_t}{t}\right|_\textrm{\textrm{step j}} = -
\frac{\gamma}{2} \left(\bdag_j\hat b_j \hrho_t - 2\hat
b_j\hrho_t\bdag_j+\hrho_t\bdag_j\hat b_j \right),
\label{master:eq} \end{equation} where
$\gamma=\rate\OmBog^2\tau^2$ and $\rate$ is the atomic arrival
rate. Hence, during step $j$ we have $\ave{\bdag_j\hat b_j}_t =
\ave{\bdag_j\hat b_j}_{0} \exp(-\gamma t)$, which vanishes at
times $t \gg 1/\gamma$, see Fig.~\ref{Fig:two-step}(a). In terms
of the original field modes, this procedure implies that the atoms
pump in phase only the two-mode squeezed state. Asymptotically,
the field state approaches \begin{equation} \hrho_\infty =
\ket{0,0}_b\bra{0,0}=\Stdag(r_\mu)\ket{0,0}_a\bra{0, 0}\St(r_\mu).
\end{equation} which is a two-mode squeezed state, whose degree of
squeezing $r_\mu$ is solely determined by the ratio
$|\Delta/\Omega|$. This state is reached independently of the
initial state of the cavity modes, provided that each step is
implemented for a sufficiently long time $T$.
\\
\indent We now analyze the proposal requirements. The two-step
procedure needs a change in the transition frequency of the two-level
atom, which can be achieved by an external static field. The
scheme needs neither atomic detection, nor control of the number
of atoms, nor of the interaction times (atomic velocities). On the
other hand, the atom must not decay during the interaction with
the cavity modes, and dissipation of the cavity field should be
negligible during the experiment. Experiments with microwave
resonators~\cite{Haroche-Colloquium,CavityQED-Walther} are
characterized by interaction times of the order of tens of $\mu$s,
which warrant negligible spontaneous decay, typically of the order
of tens of ms. The time $T$ required for each step to reach
$\langle\bdag_j\hat b_j\rangle_T\sim\bar{n}_\infty$ depends on the
initial value $\langle\bdag_j\hat b_j\rangle_0=:\bar{n}_0$ through
$T = \gamma^{-1}|log(\bar{n}_\infty/\bar{n}_0)|$.
Fig.~\ref{Fig:two-step}(b) displays the estimated total
experimental times and corresponding average number of photons per
mode at steady state as a function of $\mu$, where
$\bar{n}_0=\mu^2/(1-\mu^2)$ when the cavity modes are in the
vacuum state at $t=0$. For the degree of squeezing
$r_\mu\approx2.1$ ($\mu=0.97$), leading to an average number of 16
photons per mode at steady state, and $\bar{n}_\infty=0.01$,
corresponding to a fidelity $F\approx 0.98$, then one has $2T\sim
19$ ms in case of an initially empty cavity ($2T\sim 22$ ms for
0.7 thermal photons). Resonators stable over this time are
available in present experiments~\cite{ultra}. Fluctuations in the
coupling with the driving field, $\delta \Omega$,  give rise to an
effective linewidth of the dressed states, and are negligible
provided that $\delta\Omega T\ll1$ during $T$. This holds if the
coherence time of the driving field is much larger than $T$, which
is easily achievable with current microwave sources. Statistical
properties of the cavity field can be evaluated by measuring the
internal states of the emerging atoms \cite{PhysRevA.49.2962}. Its state can also be determined by reconstructing
the corresponding Wigner function, by suitably generalizing the
schemes proposed in~\cite{lougovski-2003-91}.  

In conclusion, a highly non-classical state of radiation can be
generated in a cavity as the steady state of the stochastic
interaction with a beam of two-level atoms.  This
final outcome does not depend on the initial state of the field.
The atoms constitute a spin reservoir, which plays no
detrimental role for quantum coherence, but it assists its
formation, leading to an EPR-like state of the electromagnetic
field. In this sense, this is an instance of reservoir engineering
within the framework of cavity quantum electrodynamics. Due to its
robustness, this proposal is an important step towards quantum
networking with atom-photon interfaces in the microwave regime.

We acknowledge discussions with M. Brune, K. Eckert, J. Eschner,
M. Hennrich, and S. Stenholm, and support by the European
commission (SCALA, Contract 015714; EMALI, MRTN-CT-2006-035369),
the Spanish MEC (Consolider Ingenio 2010 "QOIT"; QLIQS,
FIS2005-08257-C02-01; Ramon-y-Cajal), the Carl-Duisberg
foundation, the Brazilian agencies CNPq, FAPERJ, FUJB, and the
Brazilian Millennium Institute for Quantum Information.


\begin{thebibliography}{99}
\bibitem{polarizations} The fields polarizations must be such that
they all couple with the dipole transition
$|g\rangle\to|e\rangle$, while coupling to other states is avoided. For Stark-shifted circular Rydberg states, 
this can be achieved by
using circular polarizations for the cavity modes and for the pump
field.
\bibitem{zero} At $\Delta=0$ no correlations between the cavity
modes are established by this interaction.
\bibitem{Footnote:Master} The same steady state can also be
achieved in the strong-coupling regime by randomizing the
interaction times.
\end{thebibliography}

\end{document}